\documentclass[12pt,preprint]{aastex}

\shorttitle{Spectroscopy of Northern NLTT Stars}
\shortauthors{Cruz \& Reid}

\begin{document}
\title{Meeting the Cool Neighbors III: Spectroscopy of Northern NLTT Stars}

\author{K. L. Cruz\altaffilmark{1}}

\affil{Department of Physics and Astronomy, University of Pennsylvania, 209 South 
33rd Street, Philadelphia, PA 19104}
\email{kelle@sas.upenn.edu}

\and

\author{I. Neill Reid\altaffilmark{1}}
\affil{Space Telescope Science Institute, 3700 San Martin Drive, Baltimore,
MD 21218; \\
Department of Physics and Astronomy, University of Pennsylvania, 209 South 33rd 
Street, Philadelphia, PA 19104}
\email{inr@stsci.edu}

\altaffiltext{1}{Visiting Astronomer, Kitt Peak National Observatory, National 
Optical Astronomy Observatory, which is operated by the Association of 
Universities for Research in Astronomy, Inc. (AURA) under cooperative agreement 
with the National Science Foundation.}

\begin{abstract} 
We present initial results of an all-sky search for late-type
dwarfs within 20 pc of the Sun using the New Luyten Two-Tenths (NLTT)
catalog cross-referenced with the 2-Micron All Sky Survey (2MASS) database.  
The results were obtained with low-resolution optical
spectroscopic follow-up of candidate nearby-stars as a preliminary test of our
methodology. M$_{J}$, derived using spectral indices, and 2MASS $J$ are used to
estimate distances.  Out of the 70 objects observed, 28 are identified as
previously unrecognized objects within 25 pc of the Sun, and up to 19 of these
are within 20 pc.  One, LP~647-~13 is an M9-type
dwarf at 10.5 pc making it one of the four closest M9 dwarfs currently known.  
We also discuss the chromospheric activity of the observed
dwarfs.

\end{abstract}


\section {Introduction} 
This is the third in a series of papers which present
the results of our survey of the low-mass residents of our immediate Solar
Neighborhood.  \citet[hereafter Paper~I]{paper1} discussed how our capabilities for
finding low-luminosity main-sequence stars has been enhanced with the
availability of the 2-Micron All Sky Survey (2MASS) \citep{2MASS}.  The 
method that we focus on in this paper is using 2MASS in conjunction with 
proper-motion catalogs --- particularly the New Luyten Two-Tenths (NLTT) catalog 
\citep{nltt}.  This strategy is one
part of a comprehensive search for previously-unrecognized nearby stars.  The
goals of the project are two-fold:  to identify late-type dwarfs within 20
pc that can be targeted for detailed study as part of the NSF/NStars
project and to use this sample to determine the mass function of low-mass
objects in the Galactic Disk.

Our first results have come from targeting high proper-motion objects from the
NLTT catalog.  As discussed in Paper I, we were able to identify a
substantial fraction of the proper-motion stars in the 2MASS database based on
location coincidence.  With this sample, we are able to select candidate nearby
dwarfs by combining the $m_{r}$ estimates from the NLTT and the near-infrared
magnitudes provided by 2MASS and using $m_r - K_s$ colors to obtain a
rough photometric distance.

As detailed in Paper~I, our initial sample of nearby-star candidates is drawn 
from NLTT objects that
have a 2MASS counterpart within a 10$\arcsec$ search radius.  While 23,795
objects were found, only 1245 have photometric properties consistent with
their being late-type dwarfs within 20 pc of the Sun.  This sample is dubbed
NLTT Sample~1.

NLTT Sample~1 has already yielded many previously unrecognized nearby
objects.  In Paper~I, we combine the 2MASS infrared magnitudes with published
optical photometry for 469 dwarfs, identifying 76 additions to the 20 pc
sample.  \citet[hereafter Paper II]{paper2} lists 48 new objects within 20 pc, five
of which are probably within 10 pc.  These were located by obtaining
optical photometry of 180 bright southern nearby-star candidates with the
facilities at the Sutherland station of the South African Astronomical
Observatory.

This paper presents the first results from spectroscopic follow-up observations
of NLTT stars.  The selection of the current sample and its overlap with the 
finalized NLTT Sample~1 are outlined in \S\ref{selection}.  Section~\ref{observations}
describes our observations.  We present spectral indices, spectral types,
absolute magnitudes, and distances for all the observed objects in
\S\ref{results}.  A discussion of our findings, particularly interesting
objects, and chromospheric activity is in \S\ref{discussion}.  
We summarize the main results in the final section.

\section{Target Selection} 
\label{selection} 
The objects presented in this paper are taken from the initial sample of 
23,795 NLTT objects that have a 2MASS counterpart within 10$\arcsec$ of 
the NLTT position and $|b| > 10\degr$, but were selected before we
finalized the criteria for defining the NLTT Sample~1.  Indeed, these
observations provided some of the basis for those criteria.

The present set of targets were required to have declinations greater than
-30\degr and right ascensions between 21$^{\mbox{h}}$ and 5$^{\mbox{h}}$.  The following color
criteria further reduced the sample to 907 objects:

\begin{eqnarray}
m_r(lim) & = & \left\{
\begin{array}{ll}
1.67(m_r - K_s) \ + 5.5, & \mbox{if } 1.5 < (m_r - K_s) \leq 3, \\
5(m_r - K_s) \ - 4.5, &  \mbox{if } 3 < (m_r - K_s) \leq 3.8, \\
1.72(m_r - K_s) \ + 8, &  \mbox{if } 3.8 < (m_r - K_s) \leq 7. \\
\end{array}
\right.
\end{eqnarray}
Objects were eliminated if $m_r > m_r(lim)$.

Primary and secondary target lists were created by invoking stricter color
criteria, designed to probe areas of color-space most likely to contain nearby,
late-type objects.  The primary list includes 52 objects which meet the above 
critera and have $J-K_s$ colors redder than 0.95 and $H-K_s$ $>$ 0.35.  The
$J-K_s$ cut eliminated 628 objects while the $H-K_s$ eliminated 818. The secondary 
list includes 119 targets, all with $R-K_s$ $>$ 5.  Taking into account the 
significant overlap between the two lists, there is a total of 127 target objects.
  Twenty-nine already have spectroscopic observations, with most identified as mid to 
late M-dwarfs (Table~\ref{known}).  Twenty-eight objects were eliminated because 
the 2MASS magnitudes were unreliable due to nearby bright stars or their
diffraction spikes, unresolved companions, or an NLTT/2MASS
mismatch.  A mismatch occurs when more than one 2MASS object is within 10$\arcsec$
of the NLTT position and the NLTT object is linked with both the
correct and incorrect 2MASS objects (see Paper I, \S3.3).  In some cases, we 
were able to correct the mismatches and observe the appropriate object.  
The resulting target list includes 70 objects --- all of which we observed and 
present here\footnote{Finder charts can be obtained from the 2MASS Survey 
Visualization and Image Server at \url{http://irsa.ipac.caltech.edu/} using 
the positions or names given in Tables~\ref{yes} and \ref{no}.}.

\begin{figure}
\figurenum{1}
\plotone{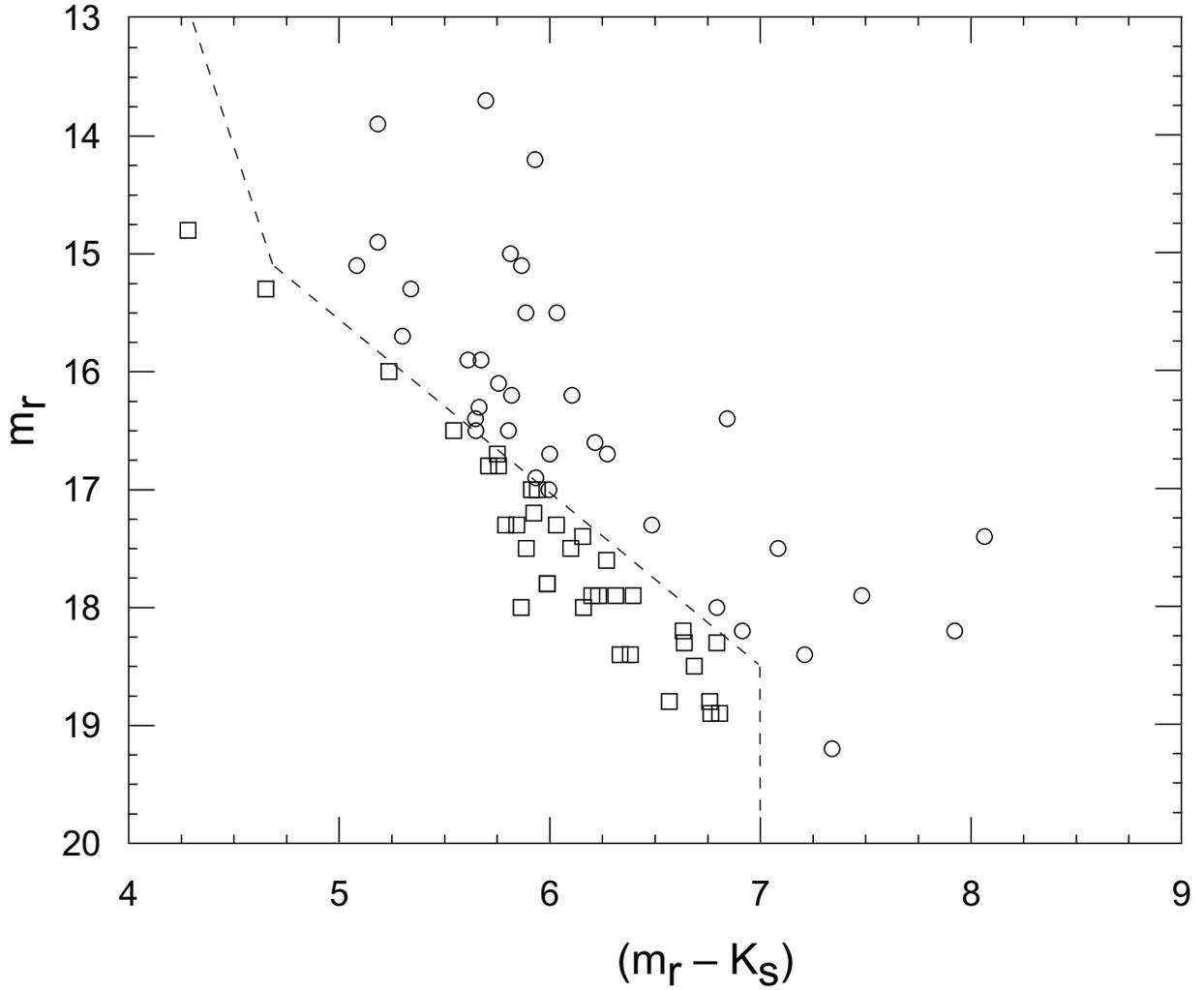}
\caption{Our 70 objects in the $(m_r,m_r-K_s)$ plane.  Circles are 
objects that are included in NLTT Sample~1 described in Paper~I.  Squares are not in the finalized sample.  The dashed line marks the limits, m$_r$(lim),
for NLTT Sample~1.}
\label{rk}
\end{figure}

Following this initial observing run, we were able to refine our color criteria
to more efficiently exclude objects beyond 20 pc.  The finalized criteria
are described in Paper~I and were used to create the NLTT Sample~1 consisting
of 1245 targets.  These observations include a significant number
of targets lying beyond the 20 parsec limit.  In Figure~\ref{rk}, we show all
of the observed objects with the finalized ($m_r,R-K_s$) color criteria
superimposed (see Paper~I, \S3.2).  Objects in the NLTT Sample~1 are listed in 
Table~\ref{yes}, while data for targets which fail to meet our final selection 
cut are presented in Table~\ref{no}.

\section{Observations} 
\label{observations} 
We obtained optical spectroscopy of our sample with the
Kitt Peak National Observatory 2.1 m telescope using the GoldCam CCD
Spectrograph.  We employed a 400 line mm$^{-1}$ grating blazed at 8000 $\mbox{\AA}$ with a
1\farcs3 slit to give a resolution of 5.1 $\mbox{\AA}$ (2.8 pixels) over the 
wavelength range 5500--9300 $\mbox{\AA}$.  We used an OG-550 blocking filter 
to block higher orders.  The observations were taken over four
nights from 2000 September 29 through October 2~(UT), all under photometric
conditions and with good seeing (between 1\arcsec and 1\farcs5).

The spectra were extracted and wavelength and flux calibrated using standard
IRAF routines. We used zero-second dark exposures taken at the beginning of each
night to remove the bias level from each exposure, via the IRAF routine CCDPROC,
which was also used to fix bad pixels. All spectra were
extracted using APALL.  Wavelength calibration was determined from HeNeAr
arcs taken after each exposure.  The spectra were flux calibrated using observations
of HD 19445 \citep{og83}, and the spectral ratios were measured using IDL scripts.

The CCD used with GoldCam suffers from fringing in the red which has an amplitude of 
$\pm3\%$ at $8000\mbox{ \AA}$, rising to $\pm10\%$ at $8400\mbox{ \AA}$.  In an attempt
to compensate for this effect, an internal-lamp flat-field exposure was taken 
after each stellar observation. However, we were unable to use these data to
correct the observed fringing in a satisfactory manner. Since the fringing does 
not affect the spectrum in the regions sampled by the measured bandstrength indices, and 
since there are no significant flat-field features shortward of $7800\mbox{ \AA}$, we have not
applied flat-field corrections to the data.

\section{Results} 
\label{results} 
The change in strength of the major features present in spectra of late-type stars
 is tied to variation in effective temperature.  Thus, we use measurements of 
the strengths of those features to estimate spectral type and 
absolute magnitude.  Band strengths can be quantified by measuring spectral 
ratios or indices.  
Table~\ref{defn} defines the spectral indices used in our study.  These are taken from 
\citet[hereafter PMSU1]{PMSU}, \citet{K99}, and \citet{ma2}, and are designed to measure 
the strengths of the most prominent features of M and early L-type
dwarfs.  The indices are calculated by taking the ratio between the summed flux 
in a region that contains an atomic or molecular feature and the summed 
flux in a nearby region that approximates the local pseudo-continuum. 
Table~\ref{indicies} lists the measurements for all of the observed objects.

\subsection{Metallicity} 
\label{diskdwarfs} 
\begin{figure} 
\figurenum{2} 
\plotone{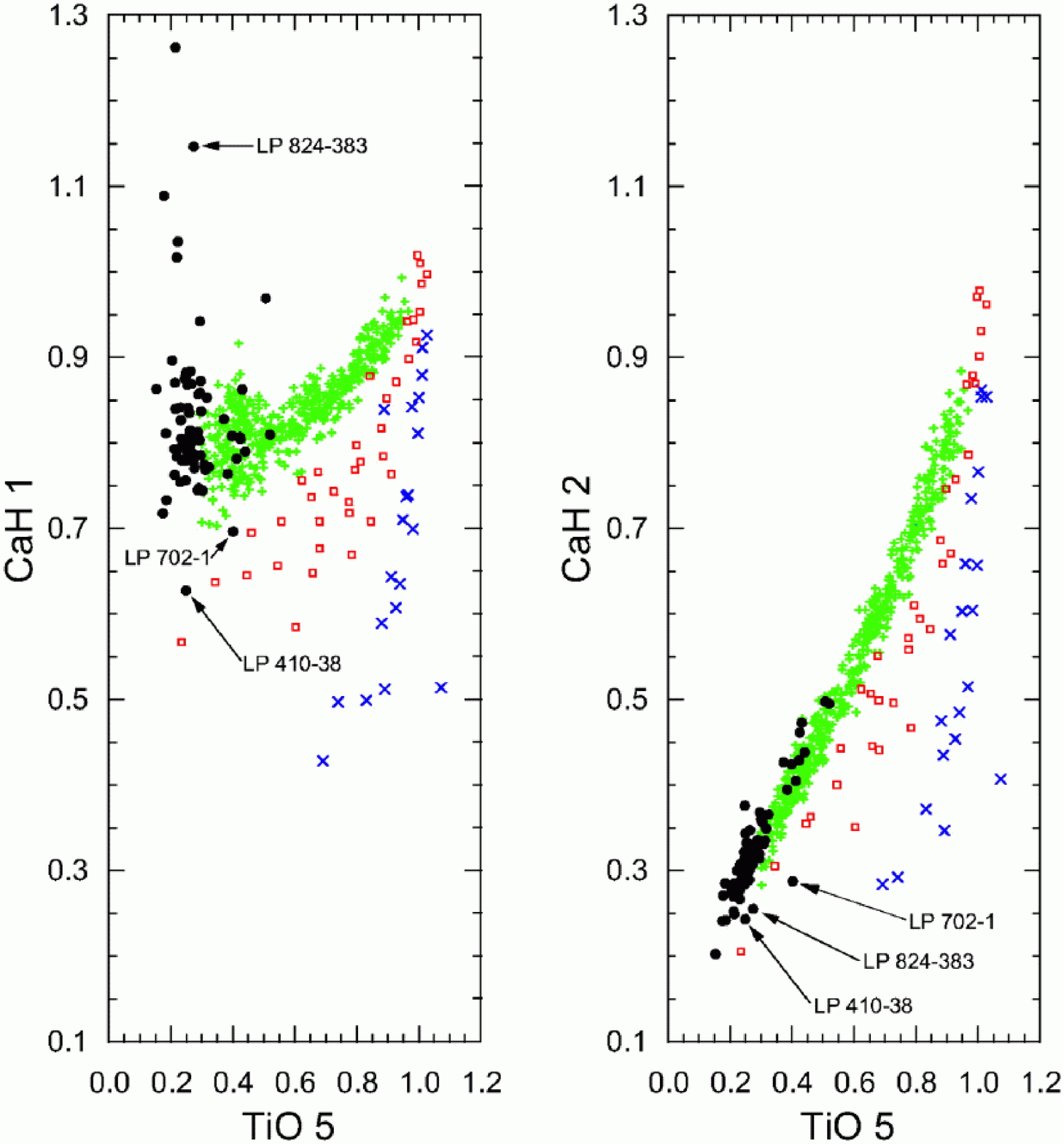}
\caption{Comparison between the CaH and TiO~5 band-strengths measured from our
observations of NLTT dwarfs and standard stars. The green plus signs mark data for
disk dwarfs from PMSU1;  open red squares are sdM subdwarfs and blue crosses are esdM
subdwarfs from \citet{subdwarfs}; our observations are plotted as solid black circles. 
Three possible M subdwarfs are identified. As discussed in the text (\S\S\ref{diskdwarfs} and \ref{sub_dwarf}),
both LP~824-383 and LP~702-1 have low signal-to-noise spectra and are probably not metal poor.}
\label{sub_dwarf_fig} 
\end{figure}

In late-type dwarfs the relative
strength of CaH and TiO absorption provides a metallicity indicator, with TiO
absorption decreasing more rapidly than CaH with decreasing metal abundance
\citep{mould}.  \citet{subdwarfs} used this behavior to define a
classification system for late-type subdwarfs, classifying stars as either
subdwarfs, sdM (intermediate abundance, [Fe/H]$\sim -1$), or extreme subdwarfs,
esdM ([Fe/H] $< -1.5$). Figure~\ref{sub_dwarf_fig} plots the CaH~1-TiO~5 and
CaH~2-TiO~5 diagrams for our sample, where data for the reference stars are
taken from PMSU1 (the disk main sequence) and \citet[sdM and esdM
sequences]{subdwarfs}. All of our targets, except LP~410-38 (2M0230) and LP~702-1 (2M2310), have
spectral indices consistent with their being near solar-abundance disk dwarfs.  
These two objects are further discussed in \S\ref{sub_dwarf}.
While the location of LP~824-383 (2M0012) in the CaH~2-TiO~5 plane is
consistent with that of an intermediate subdwarf, the spectrum has a low signal-to-noise ratio and the
CaH~1 and CaH~2 measurements are not reliable. 

\subsection{Spectral Types} 
\label{spectraltype} 
\begin{figure}
\figurenum{3}
\plotone{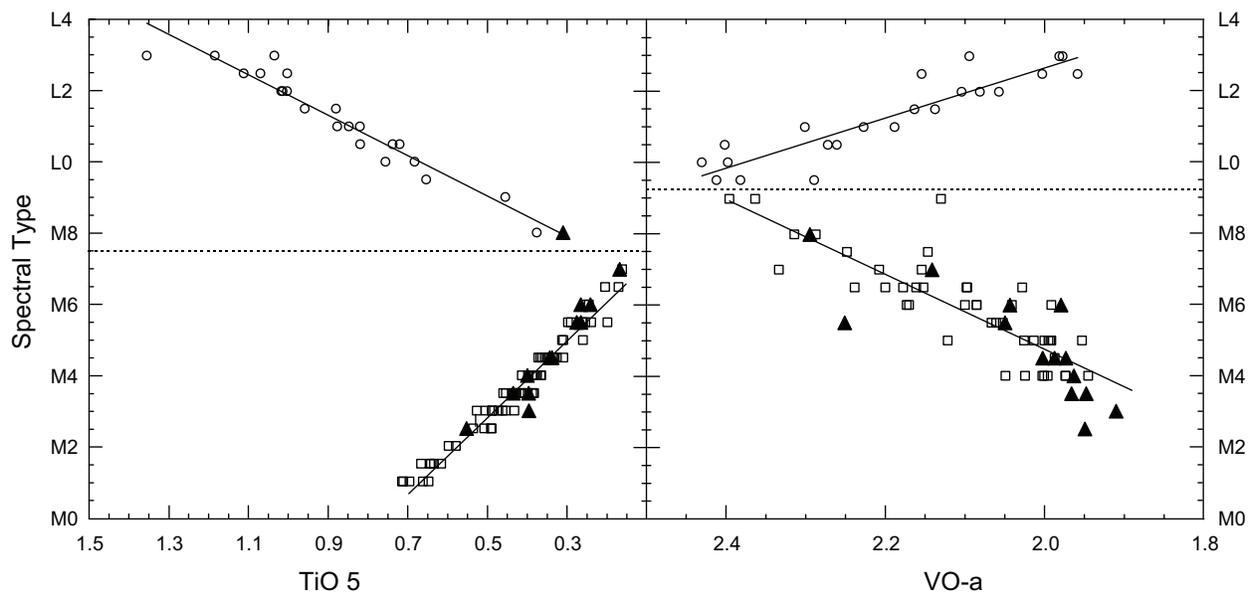}
\caption{TiO~5 and VO-a spectral type calibrations.  Late-type calibrating
objects (later than M7 for TiO~5 and M9 for VO-a) are plotted as open circles,
earlier types as open squares, and our standards as
solid triangles.  The dotted line illustrates the separation of the two trends.
Data are from PMSU1 and \citet{67}.  In addition, our standards were included 
in calculating the late-type TiO~5 and early-type VO-a relations. The early 
TiO~5 relation was adopted from PMSU1 and our standards are overplotted to 
show their agreement.}
\label{spec_cal}
\end{figure}

We have defined the spectral
type calibration using data for nearby stars and brown dwarfs with published
spectral types (\citet{67}; PMSU1).  We have supplemented our own observations
with Keck Low-Resolution Infrared Spectrometer \citep[LRIS]{LRIS} spectra of
late-M and L dwarfs obtained by INR and collaborators as part of the 2MASS Rare
Object Project\footnote{Most of the spectra are publicly available from
\url{http://dept.physics.upenn.edu/$^{\sim}$inr/}}.  Spectral ratios for the 
standards were measured using the same scripts used for the KPNO data presented 
here.  The indices that best correlate with spectral type are TiO~5 and VO-a.  
Both indices are double-valued, with TiO~5 reversing in strength at M7 and 
VO-a at M9.  For the early TiO~5 sequence, we adopt the relation found by PMSU1.  
The data and the calibration curves are plotted in Figure~\ref{spec_cal}.
\clearpage

The spectral type calibration relations are:

\begin{displaymath}
S_{p} =-10.775\mbox{(TiO 5)} + 8.200, \quad \mbox{(TiO 5)} \leq 0.75, \quad \sigma = 0.5\mbox{ subclasses},
\end{displaymath}
\begin{displaymath}
S_{p} = 5.673\mbox{(TiO 5)} + 6.221, \quad \mbox{(TiO 5)} \geq 0.3, \quad \sigma = 0.38\mbox{ subclasses, 23 stars},
\end{displaymath}
\begin{displaymath}
S_{p} = 10.511\mbox{(VO-a)} - 16.272, \quad \sigma = 0.82\mbox{ subclasses, 59 stars},
\end{displaymath}
\begin{displaymath}
S_{p} = -7.037\mbox{(VO-a)} + 26.737, \quad \sigma = 0.50\mbox{ subclasses, 22 stars}.
\end{displaymath}
In principle, these relations yield up to four estimates of the spectral type.
  However, the fact that the two indices' trends reverse at different spectral
 types allows us to resolve the ambiguity since only one pair of solutions 
agree.  We take the spectral type to be the weighted average of the results 
(one from TiO 5 and one from VO-a) rounded to the nearest half spectral type.  
The resulting uncertainty is $\pm$0.5 subclasses.


\subsection{Absolute Magnitudes and Derived Distances} 
\label{mjandd} 
\begin{figure}
\figurenum{4}
\plotone{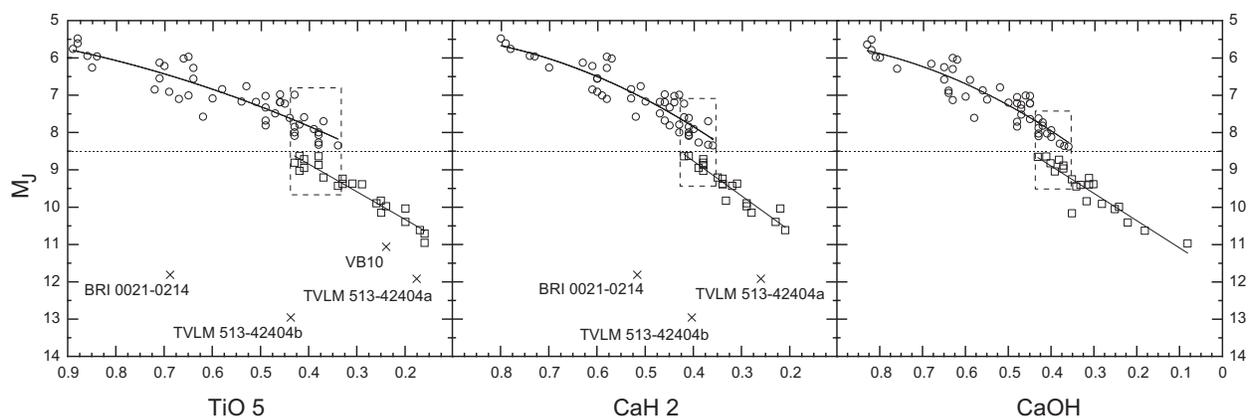}
\caption{Absolute magnitude calibration using K7 and later dwarfs.  Circles are the data included in the calibration for brighter objects while the
fainter calibration objects are plotted as squares.  The dashed boxed
regions ($0.34~\leq$~TiO~5~$\leq~0.43$, $0.36~\leq$~CaH~2~$\leq~0.42$, 
$0.36~\leq$~CaOH~$\leq~0.43$) show where the calibration is double-valued.  The 
crosses, VB~10~(M8), TVLM~513-42404a~(M7), TVLM~513-42404b~(M9), and 
BRI~0021-0214~(M9.5), show that the trends reverse for later types.}
\label{mag_cal}
\end{figure}

The absolute magnitude/band strength calibration was defined using a sample of 68
late-type dwarfs (from K5 to M7) with well-determined trigonometric parallaxes,
taken from the nearby stars surveyed by PMSU1.  The latter authors
provide band strength measurements for a variety of indices.  We find that
color-magnitude diagrams using the TiO~5, CaH~2, and CaOH indices show the
smallest dispersion in the main sequence, and hence the best prediction of
absolute magnitude.  The calibrating data and curves are shown in
Figure~\ref{mag_cal}.

There is a jump in all three color-magnitude diagrams at M$_J$ = 8.5.  This
break is discussed in detail in \S4.2 of Paper I.  To accommodate this jump, we
have fit the main sequence in two separate regions.  The absolute magnitude
calibration relations are:

\begin{displaymath} 
M_{J} = 2.79\mbox{(TiO 5)}^{2} - 7.75\mbox{(TiO 5)}+10.49,\quad \mbox{(TiO 5)} \geq 0.34, \quad \sigma = 0.35 \mbox{ mag., 46 stars},
\end{displaymath}
\begin{displaymath}
M_{J} = -7.43\mbox{(TiO 5)} + 11.82,\quad \mbox{(TiO 5)} \leq 0.43, \quad \sigma = 0.19 \mbox{ mag., 22 stars},
\end{displaymath}
\begin{displaymath}
M_{J} = 6.50\mbox{(CaH 2)}^{2} - 13.24\mbox{(CaH 2)} + 12.10, \quad \mbox{(CaH 2)} \geq 0.36,\quad \sigma = 0.33 \mbox{ mag., 46 stars},
\end{displaymath}
\begin{displaymath}
M_{J} = -9.33\mbox{(CaH 2)} + 12.50, \quad \mbox{(CaH 2)} \leq 0.42, \quad \sigma = 0.19 \mbox{ mag., 20 stars},
\end{displaymath}
\begin{displaymath}
M_{J} = 5.67\mbox{(CaOH)}^{2} - 11.99\mbox{(CaOH)} +11.82, \quad \mbox{(CaOH)} \geq 0.36,\quad \sigma = 0.32 \mbox{ mag., 46 stars},
\end{displaymath}
\begin{displaymath}
M_{J} = -7.31\mbox{(CaOH)} + 11.80, \quad \mbox{(CaOH)} \leq 0.43, \quad \sigma = 0.28 \mbox{ mag., 21 stars}.
\end{displaymath}

As with the spectral type calibration, there is a region of overlap
between the relations for each spectral index where $M_J$ is ambiguous.  
These regions ($0.34~\leq$~TiO 5~$\leq~0.43$, $0.36~\leq$~CaH~2~$\leq~0.42$, 
$0.36~\leq$~CaOH~$\leq~0.43$) are enclosed by dashed boxes in Figure~\ref{mag_cal}.  
The calibration objects that lie in these regions are mostly M3.5-, M4-, and
M4.5-type dwarfs.  The measured spectral ratios give an unambiguous estimate of 
$M_J$ for most of the NLTT dwarfs.  However, in a few cases, one or all of the predictions 
for $M_J$ are multi-valued (there are no cases where two of the three indices 
give ambiguous predictions).  In cases where only one index yields two values 
for $M_J$, we adopt the value that is the closest to the predictions from the 
other two spectral ratios.  The value is included in the weighted average of 
$M_J$ if it lies within 3$\sigma$ of either of the other two values.  In the 
case of all three indices yielding two values for $M_J$, we quote all six 
values and compute two weighted averages using the sets of predictions that 
are closest to each other.  We therefore give two estimates of $M_J$ and thus, two
estimates for the distance for a small number of stars in Tables~\ref{yes} and 
\ref{no} (e.g., G~75-35 (2M0241) in Table~\ref{yes}).  Reliable distances for these objects 
can only be derived by obtaining trigonometric parallaxes.

The uncertainty in $M_J$ was calculated by adding in quadrature two
contributions to the uncertainty: the RMS of the weighted average based on
the RMS of the individual calibration fits stated above and the standard
deviation of the values of $M_J$ given by the different spectral ratios.  
Using this scheme and the 2MASS apparent $J$ magnitude, we estimate $M_J$ and 
the distance to all of our (disk dwarf) targets (Tables~\ref{yes} and~\ref{no}).

\section{Discussion} 
\label{discussion}

\subsection{Additions to the Nearby-Star Sample}
Our spectroscopic observations confirm that combining optical and near-infrared
photometry is an effective means of identifying new stellar neighbors, even when
the optical photometry is as unreliable as the magnitudes listed in the NLTT catalog.
In particular, of the 35 stars listed in Table~\ref{yes}, selected on the basis of our
finalized color-magnitude criteria, up to 19 (54\%) are likely to lie within
20 pc of the Sun, while up to 27 (77\%) probably lie within the 25 pc
sample.  Several stars require particular comment. 

\subsubsection{LP 647-13}
\begin{figure}
\figurenum{5}
\plotone{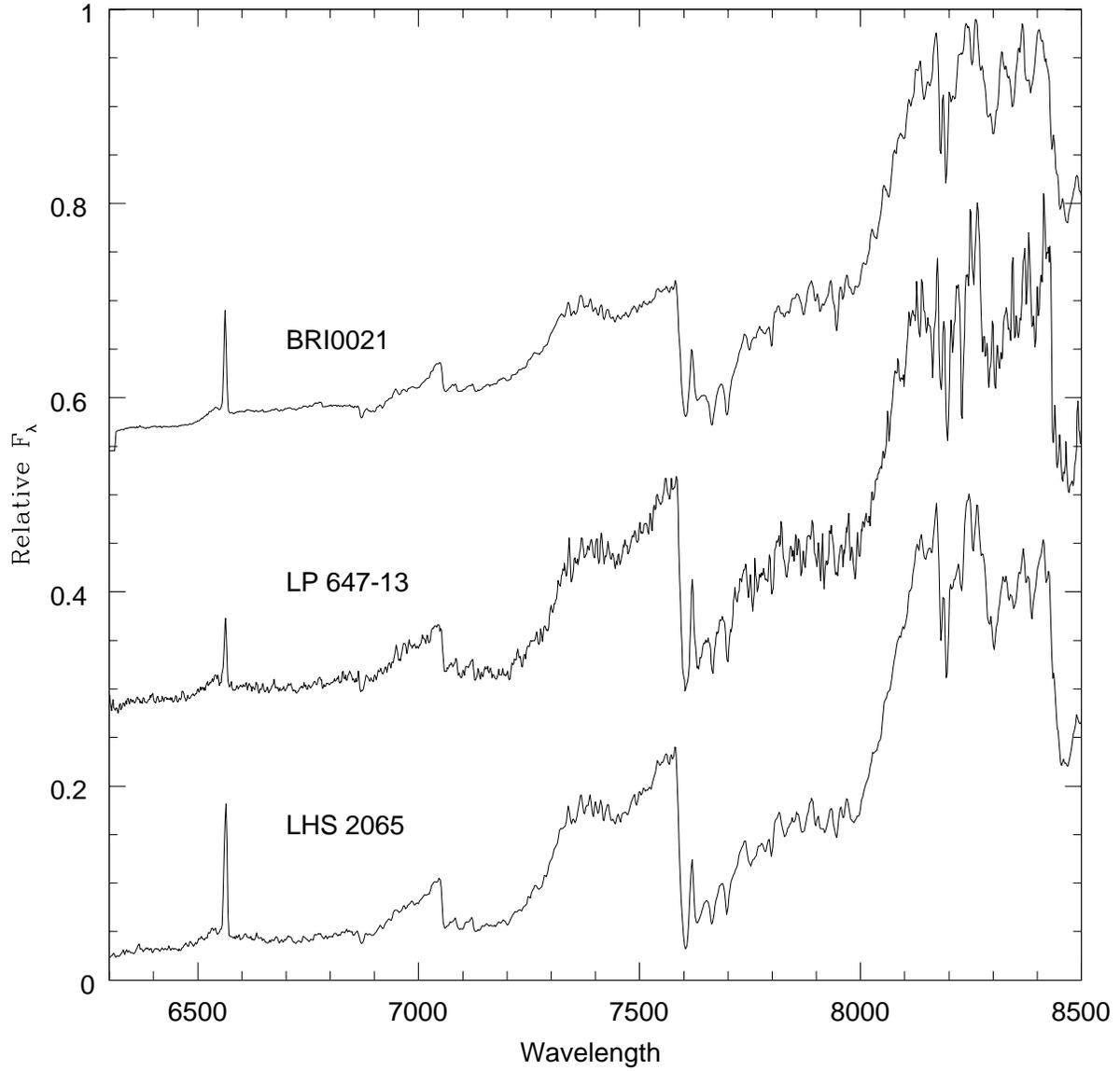}
\caption{Spectrum of LP~647-13, an M9 dwarf at 10.5 pc.  Keck LRIS spectra of the
M9 and M9.5 standards, LHS~2065 and BRI~0021-0214, are shown for comparison.}
\label{M9}
\end{figure}

At spectral type M9, LP~647-13 (2M0109) is the latest of the
NLTT dwarfs in the present sample, falling beyond the range of validity of the
absolute magnitude calibrations plotted in Figure~\ref{mag_cal}.  Figure~\ref{M9} compares our spectrum of this object with Keck LRIS data for
the M9 standard LHS~2065 and the M9.5 standard BRI~0021-0214.  There are obvious
strong similarities between LP~647-13 and LHS~2065. \citet{K95} list absolute
magnitudes for these two standard stars and for two other M9 dwarfs in the
immediate Solar Neighborhood: $M_K=10.33$ for LHS~2065; $M_K=10.22$ for
BRI~0021; $M_K=10.46$ for LHS~2924; and $M_K=10.24$ for TVLM~868-110638.  A
straight average gives $M_K=10.31\pm0.11$ magnitudes.  Applying that value
gives a distance of only 10.5 pc to LP~647-13, making it one of the four
closest M9 dwarfs currently known, along with LHS~2065 at 8.5 pc, LHS~2924 at 10.5 pc, and DENIS-P J104814.7$-$395606.1 at $\sim$5 pc \citep{delfosse}.  

\subsubsection{LP 763-38}
\begin{figure}
\figurenum{6}
\plotone{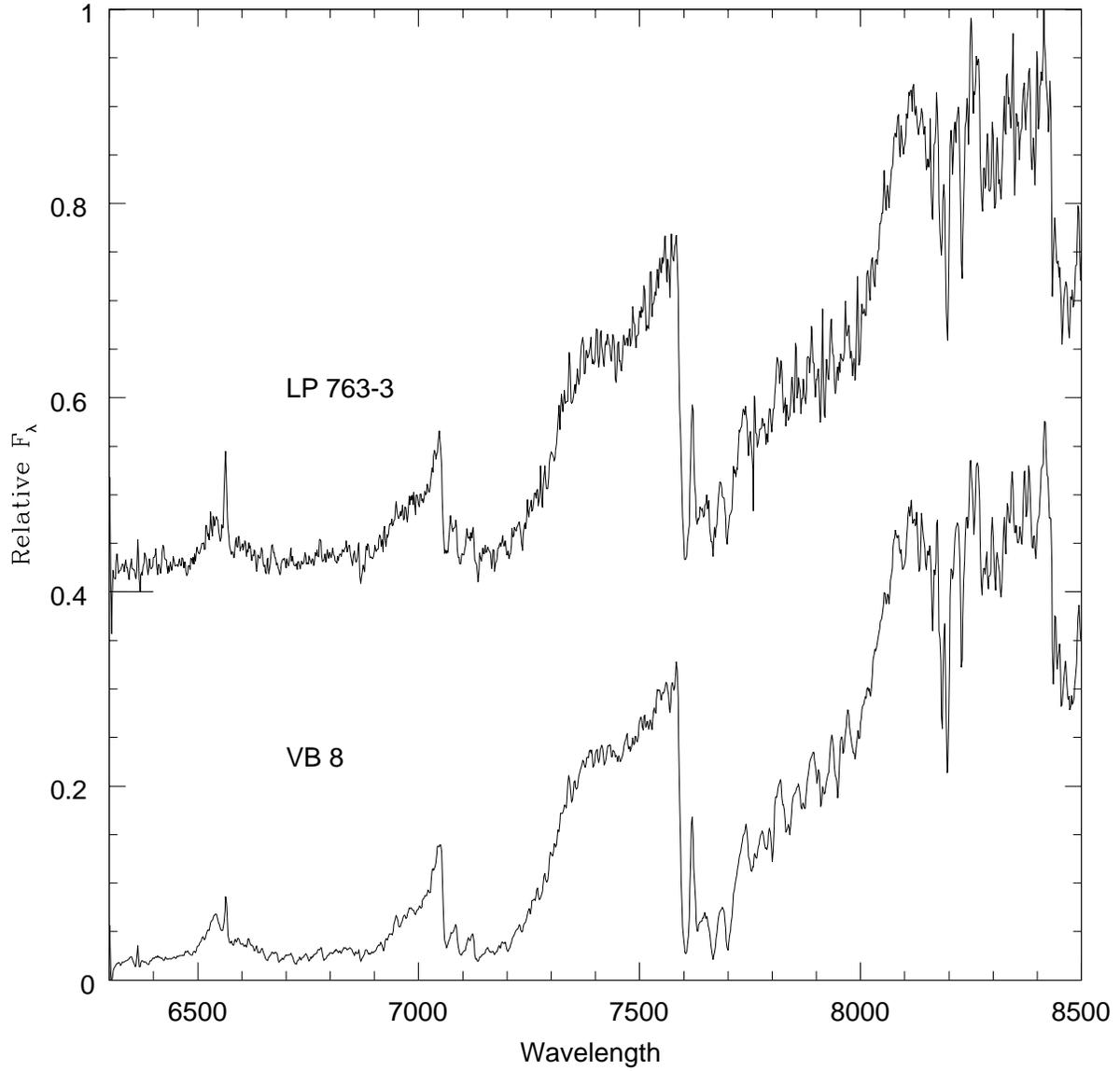}
\caption{Spectrum of the M7 dwarf, LP~763-3. The spectrum of the M7 standard, VB~8, is shown
for comparison.}
\label{M7}
\end{figure}

This dwarf (2M2337) has spectral indices which place it at the
extreme limit of validity of our calibration.  Figure~\ref{M7} plots
our spectral data and compares those with the M7 standard, VB~8 (Gl 644C). 
Given the strong similarities, we classify LP 763-38 as spectral type M7,
and estimate the distance using VB~8 ($M_K$=9.76) as a template.  Matching
that value against the observed $K$ magnitude of 11.206 for LP 763-38
gives a distance modulus of 1.47 magnitudes, or a distance estimate
of $20.0\pm3.0$ pc. 

\subsection{Possible Subdwarfs: LP~410-38 \& LP~702-1}
\label{sub_dwarf}
\begin{figure}
\figurenum{7}
\plotone{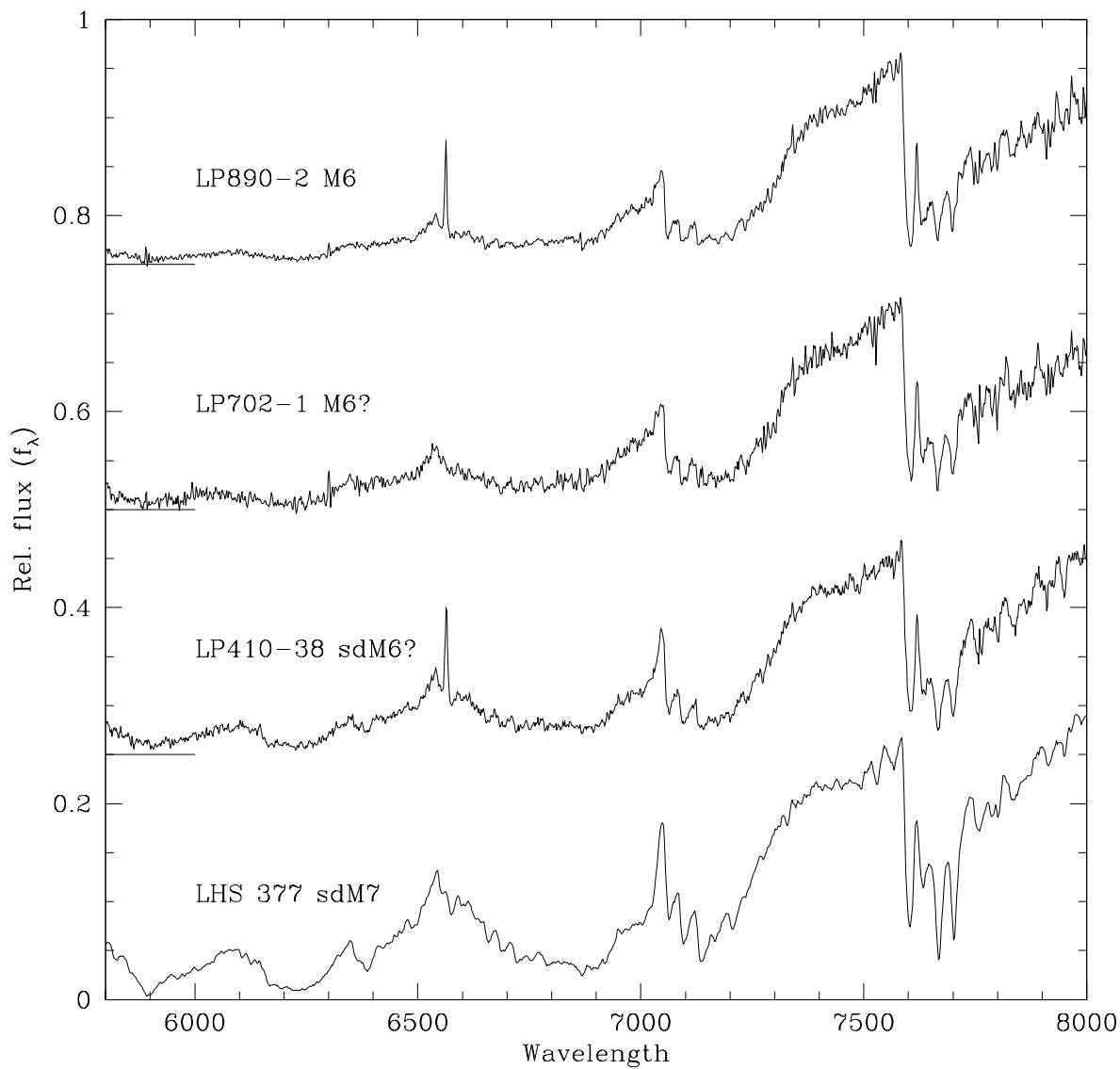}
\caption{A comparison between our spectra of the two subdwarf candidates, LP~702-1 and
LP~410-38, and data for an M6 dwarf (LP~890-2) and the late-type intermediate subdwarf, 
LHS~377 (a Keck LRIS spectrum). 
As discussed in the text, while LP~702-1 is probably misclassified due to
the reatively low signal-to-noise ratio, LP~410-38 shows the enhanced hydride absorption
characteristic of mildly metal-poor subdwarfs. }
\label{plsub}
\end{figure}

As discussed above, our band-strength meaurements
suggest that these two stars (2M0230 and 2M2310, in Table 3) are intermediate-abundance 
subdwarfs. Figure~\ref{plsub} compares our spectra against data for
LP~890-2 (2M0413, Table 2), an M6 dwarf in our NLTT sample, and 
LHS~377, one of the coolest-known intermediate subdwarf (sdM7, \citet{subdwarfs}). 
LP~702-1 is clearly similar to LP~890-2, suggesting that the 
subdwarf-like spectral indices may reflect the relatively low signal-to-noise ratio of
our spectrum.  LP~410-38, on the other hand, has spectral characteristics
which are closer to LHS~377, notably the enhanced CaH absorption at 6400 and 7000\mbox{ \AA}.
We therefore classify LP~702-1 as an M6, near-solar abundance disk dwarf, but 
identify LP~410-38 as an intermediate subdwarf, spectral type sdM6.  We adopt 
$M_J=10.15\pm0.16$ for LP~702-1.  This was computed by averaging 
the values of $M_J$ for all of the M6-type dwarfs in our sample.  This yields a
distance estimate of $37.0\pm5.0$ pc.  The \citet{subdwarfs}
subdwarf sample does not include any sdM6 stars, but both LHS~377 and LHS~407 (sdM5)
have measured parallaxes (\citet{monet} and \citet{ra},
respectively). \citet{ra} also present JHK photometry for LHS~407, while
\citet{l00} list such data for LHS~377.  Combining those measurements
gives $M_K=9.74\pm0.4$ for LHS~377 and $M_K=9.55\pm0.8$ for LHS~407.
We therefore adopt $M_K=9.7$ for LP~410-38, giving a distance estimate of 
$18.0\pm5.0$ pc.
We note that the H$\alpha$ emission evident in LP~410-38 is unusual, but not
unprecedented, in late-type subdwarfs, and might reflect the presence of a
close companion, as with Gl~455 and Gl~781 \citep{subdwarfs}.

\subsection{Chromospheric Activity}
\label{activity}

Chromospheric activity, as evidenced by emission at either the Ca {\small{\rm{II}}} H \& K 
lines or the Balmer series, is common among late-type dwarfs.  A significant 
number of the NLTT dwarfs exhibit H$\alpha$ emission, as evidenced by H$\alpha$ 
indices exceeding 1.0 in Table~\ref{indicies}.  We have used the options 
available in the IRAF routine SPLOT to measure equivalent widths and line 
fluxes for 43 stars, and the results are listed in Table~\ref{halpha}. 
Our observations set a typical upper limit of 0.75\mbox{ \AA} on H$\alpha$ 
emission in the remaining stars.  This fraction of $\sim60\%$ 
is broadly consistent with the expected proportion of dMe dwarfs at spectral 
types of M5 to M6 (see Figure 6 in \citet{NN}).  

Equivalent width is still often used to 
characterize the level of activity but, as pointed out originally by \citet{rhm} 
and later by \citet{basri}, 
this approach fails to take into account the decreased continuum level 
in later-type stars.  A more effective means of gauging the relative activity 
of dwarfs spanning a wide range of spectral types (temperatures) is to consider 
the fraction of the total flux emitted as line emission, specifically 
$F_\alpha / F_{bol}$, where F$_\alpha$ is the total flux in the H$\alpha$ line.

\begin{figure}
\figurenum{8}
\plotone{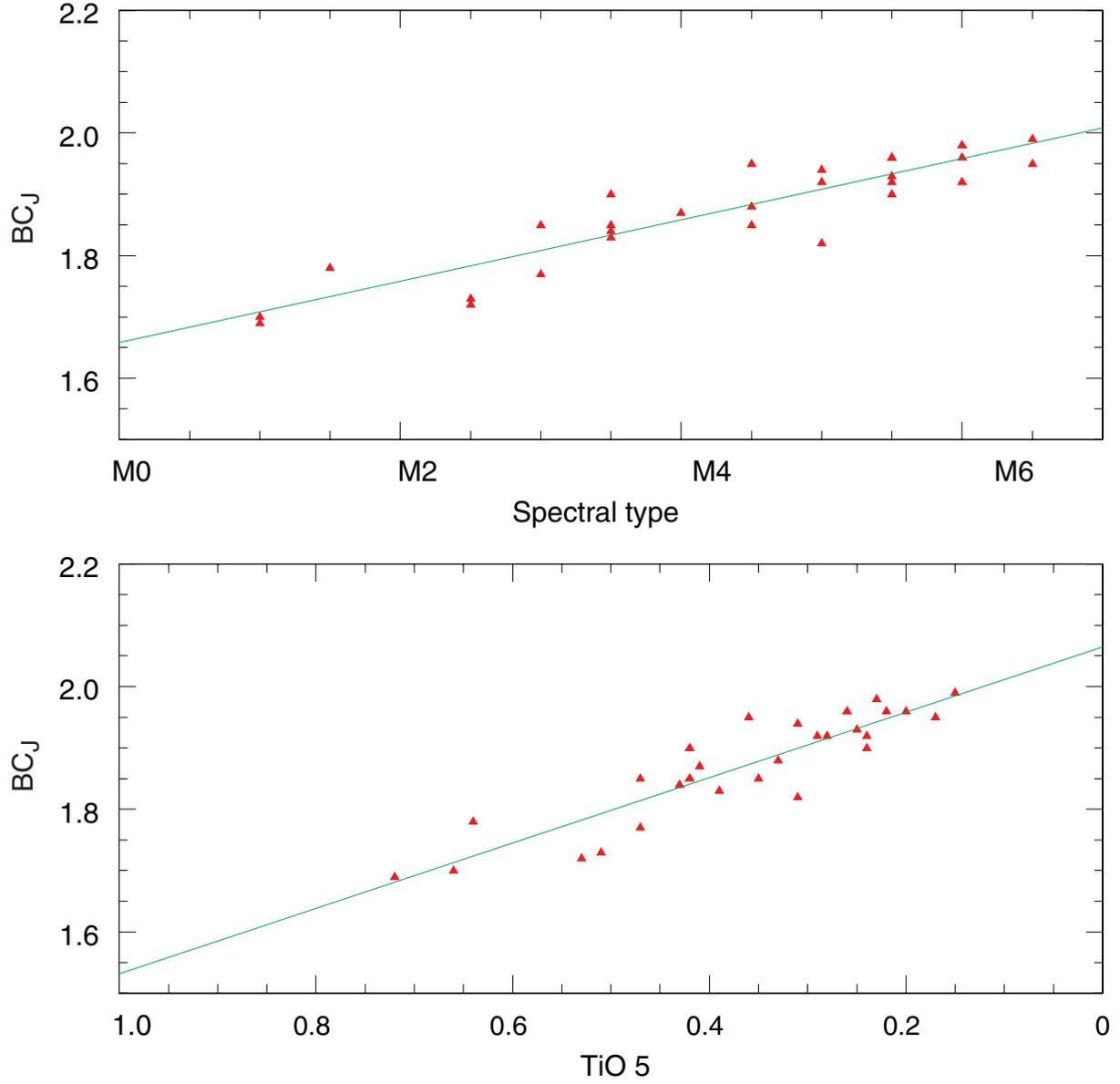}
\caption{$J$-band bolometric correction as a function of spectral type and 
TiO~5 index for M0 to M7 dwarfs based on data from \citet{l00}.}
\label{bcj}
\end{figure}

Our spectra give a direct measure of F$_\alpha$. In order to determine 
F$_{bol}$, we need to estimate bolometric corrections for the NLTT dwarfs.  We 
can calculate the latter using data from \citet{l00} observations of 28 nearby 
M~dwarfs with spectral types between M1 and M6.5.  Figure~\ref{bcj} plots the 
$J$-band bolometric corrections for those stars as a function of both spectral type 
and TiO~5 index, taking the latter from PMSU1.  Both correlations are well described by linear relations
\begin{displaymath}
BC_J \ = \ (1.658\pm0.021) \ + \ (0.050\pm0.005) \times {\rm sp. \ type}, \quad \sigma=0.036,
\end{displaymath}
and
\begin{displaymath}
BC_J \ = \ (2.065\pm0.020) \ - \ (0.533\pm0.050) \times {\rm TiO~5}, \quad \sigma=0.037.
\end{displaymath}
We have used the latter relation to estimate bolometric corrections for the 
NLTT dwarfs with spectral types M1 to M6.5; we adopt BC$_J$=1.9 magnitudes 
for later spectral types \citep{r01}.  Table~\ref{halpha} lists the resulting values of 
$\log{F_\alpha / F_{bol}}$ for dwarfs with measurable H$\alpha$ emission.

\begin{figure}
\figurenum{9}
\plotone{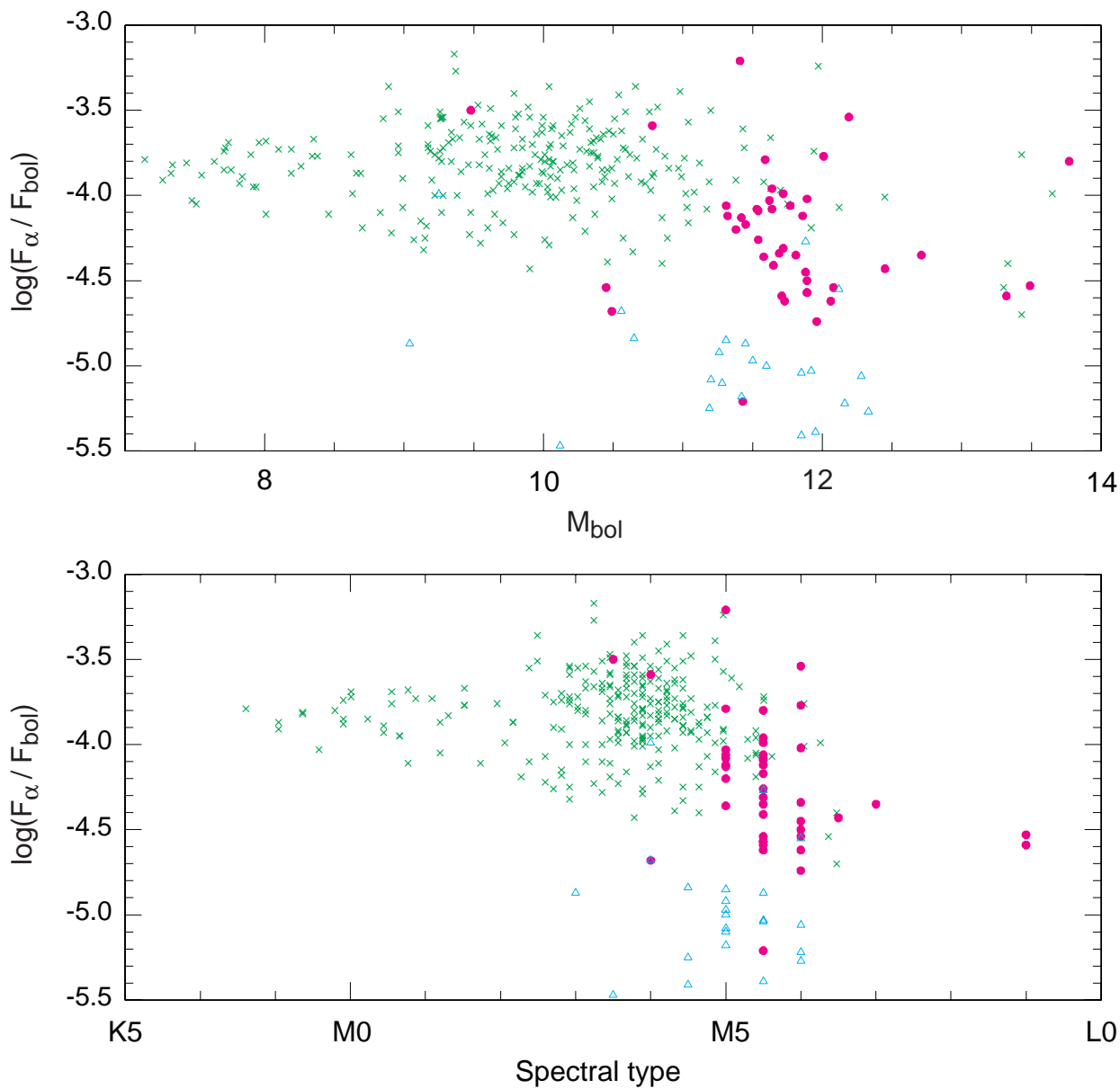}
\caption{Distribution of chromospheric activity amongst the NLTT M dwarf sample, plotted against
M$_{bol}$ (\emph{top}) and spectral type (\emph{bottom}). 
Data for nearby dMe stars from the PMSU2 sample are plotted as green crosses; NLTT 
dwarfs with detected H$\alpha$ emission in the present sample are plotted as solid magenta circles; 
open blue triangles mark upper limits for the remaining NLTT dwarfs. }
\label{ha}
\end{figure}

Figure~\ref{ha} compares the distribution of activity amongst the present sample against 
data for nearby emission-line M dwarfs from \citet[hereafter PMSU2]{PMSU2}.  2M0203 (LP~352-79) 
stands out as the most active star in the sample, with ${\log{F_\alpha / F_{bol}}~=~-3.21}$.  
We also note that \citet{NN} failed to detect H$\alpha$ emission in the M9 dwarf, 2M0350+1818, 
while we measure an H$\alpha$ equivalent width of 13.3\mbox{ \AA}.  Previous observations have 
shown that moderate-strength flares tend to occur with a duty cycle of a few percent amongst 
ultracool (spectral types later than M6.5) dwarfs \citep{r99b,ma}, and this mechanism probably accounts 
for the relatively high levels of activity in both these dwarfs.

Considering the overall distribution in Figure~\ref{ha}, the (mainly) M5/M6 dwarfs observed 
in this paper are clearly less active, on average, than the dMe dwarfs in the PMSU sample.  This 
is not unexpected, since recent studies indicate that the average level of activity is 
significantly lower amongst chromospherically-active ultracool dwarfs \citep{NN,basri01}.  Indeed, 
our observations bridge the gap between the ultacool datasets and the PMSU stars, which include 
few dwarfs between spectral types M5 and M7.  Our data show that the average level of activity 
amongst dMe dwarfs falls from 
$\langle \log{F_\alpha / F_{bol}} \rangle = -3.9$ at spectral types earlier than M5 to 
$\langle \log{F_\alpha / F_{bol}} \rangle \sim -4.25$ for M5.5 dwarfs.  Activity declines 
even further at later types, with 
$\langle \log{F_\alpha / F_{bol}} \rangle \sim -5$ at spectral type M9 and only 10 to 20\% 
of early-type L dwarfs having detectable H$\alpha$ emission.

\section{Summary} 
\label{summary} 
We have presented spectroscopic observations of 70 late-type dwarfs selected from the
NLTT proper motion catalog as probable members of the immediate Solar Neighborhood
based on their optical/near-infrared photometric properties.  Of these 70 objects, 28 are 
found to be previously unrecognized stars within 25 pc of the Sun; 13 lie within
20 pc. 

In addition to identifying a small sample of new members of the local stellar community, 
the observations described in this paper lay the foundations for the analysis
of future observations.  We have identified and calibrated a number of narrowband
spectral indices which can be used to determine spectrophotometric parallaxes and
spectral types for M~dwarfs.  Based on those calibrations, we have refined the
photometric selection criteria used to identify candidate nearby stars from our
cross-referencing of the NLTT catalog against the 2MASS database.  Future papers
will apply the techniques outlined in this paper to spectroscopic observations of
a larger sample of nearby-star candidates.

\acknowledgements 
This research was supported partially by a grant from the
NASA/NSF NStars initiative, administered by JPL,
Pasadena, CA.  KLC acknowledges support from a NSF
Graduate Research Fellowship. This research has made use of the SIMBAD
database, operated at CDS, Strasbourg, France and
the NASA/IPAC Infrared Science Archive, which is operated by the JPL, California Institute of Technology, under contract with NASA. We would also like to thank our
telescope operators at KPNO, Bill Gillespie and Hillary Mathis. This
publication makes use of data products from the Two Micron All Sky Survey,
which is a joint project of the University of Massachusetts and the Infrared
Processing and Analysis Center/California Institute of Technology, funded by NASA and the NSF.






\clearpage
\begin{deluxetable}{ccccrccrrrrccccc}
\tabletypesize{\tiny}
\rotate
\tablewidth{0pt}
\tablecolumns{16}
\tablecaption{Previously Known Objects in Our Sample of NLTT Targets.} 

\tablehead{
\colhead{LP/NLTT}   &  \colhead{Other}  &  \colhead{$\alpha$} &   \colhead{$\delta$} &     \colhead{M$_V$}   &   \colhead{Ref$_V$} & \colhead{$M_J$} &  \colhead{$m_r$} &  \colhead{$J$}    &  \colhead{$H$}   &  \colhead{$K_s$} &  \colhead{d}     & \colhead{Ref$_{\pi}$} & \colhead{Ref$_{ph}$} & \colhead{Sp.}  & \colhead{Ref$_{sp}$}\\
\colhead{Name}      &  \colhead{Name}   &  \colhead{(J2000)}  &   \colhead{}         &     \colhead{}        &   \colhead{}        & \colhead{}      &  \colhead{}      &  \colhead{}     &  \colhead{}    &  \colhead{}      &  \colhead{(pc)}  & \colhead{}            &  \colhead{}          & \colhead{Type} &  \colhead{}}
\startdata
292- 67 &            LHS 112 &        00 20 29.22 & \phs33 05 08.28  &  15.57 &  2 &$ 9.81\pm0.10$ &    15.2 & 10.312  &  9.734   &  9.347   &$ 12.6\pm0.6  $ & 1 &       & M5.5  & 2 \\
349- 25 &              &              00 27 55.91 & \phs22 19 32.89  &\nodata &    &$10.99\pm0.36$ &    17.0 & 10.608  &  9.970   &  9.561   &$  8.5\pm1.4  $ &   & 3     & M8.0  & 3 \\
766- 87\tablenotemark{a} & LHS 1146 & 00 46 20.44 & $-$19 24 43.95   &\nodata &    &$10.30\pm0.17$ &    18.2 & 12.694  & 12.094   & 11.711   &$ 30.2\pm2.3  $ &   & 4     & M6.0  & 4 \\
BD+22 176B\tablenotemark{b}&GL 53.1B& 01 07 38.52 &  \phs22 57 21.91 &  12.01 & 12 &$ 7.90\pm0.05$ &  \nodata&  9.494  &  8.899   &  8.693   &$ 20.8\pm0.5  $ & 5 &       & M3.0  & 6 \\
768- 26 &            GL 65B &         01 39 01.21 & $-$17 57 02.62   &  15.60 &  6 &$ 9.21\pm0.03$ &    12.2 &  6.302  &  5.685   &  5.358   &$  2.62\pm0.04$ & 7 &       & M5.5  & 6 \\
768- 27 &            GL 65A &         01 39 01.21 & $-$17 57 02.62   &  15.47 &  6 &$ 9.21\pm0.03$ &    12.7 &  6.302  &  5.685   &  5.358   &$  2.62\pm0.04$ & 7 &       & M5.5  & 6 \\
468-199\tablenotemark{a}&LHS 1294 &   01 45 45.29 &  \phs13 06 00.39 &\nodata &    &$ 6.65\pm0.16$ &    18.0 & 12.783  & 12.088   & 11.755   &$168.9\pm12.2$  &   & 4     & M1.5  & 4 \\
30- 55 &             G 245-040 &      02 01 54.02 &  \phs73 32 32.13 &  13.87 &  2 &$ 9.11\pm0.68$ &    14.1 &  9.252  &  8.669   &  8.382   &$ 11.2\pm3.4  $ &   & 2     & M4.5  & 2 \\
\enddata
\tablenotetext{a}{Used TiO~5 and CaH~2 indices and our calibrations to find $M_J$ and distance.}
\tablenotetext{b}{Distance based on trigonometric parallax for companion, GL 53.1A}
\tablenotetext{c}{The system made up of two NLTT objects, LP~469-118 and its close companion G 035-027, are unresolved by 2MASS and the photometry is not reliable.}
\tablenotetext{d}{Used TiO~5 and CaH~2 indices and our calibrations to find $M_J$ and a distance estimate.  Final estimate is average of this value with estimates based on optical photometry}
\tablecomments{
Column 1 lists the designation from the NLTT catalog.  Identifiers without prefixes are Luyten-Palomar (LP) Survey numbers.  Other identifiers are R=Ross and BD=Bonner Durchmusterung. 
Column 2 lists alternate designations.  
Columns 3 and 4 list the position of the 2MASS source.
Column 5 lists $M_V$ and Column 6 gives the source.
Column 7 gives the value of $M_J$ obtained using our distance estimates.  
Column 8 lists the $m_r$ given in the NLTT.  
Columns 9-11 list the infrared photometry from 2MASS. 
Column 12 lists the distance estimates, and Column 13 and 14 give the source of the trigonometric parallax or photometry upon which the distance estimate is based.
Column 15 lists the spectral type and Column 16 gives the source.
}
\tablerefs{(1)~\citet{usno85}; (2)~PMSU1; (3)~\citet{NN}; (4)~\citet{LHS}; (5)~\citet{hip}; (6)~\citet{gliese}; (7)~\citet{usno80};
 (8)~\citet{monet}; (9)~\citet{dahn}; (10)~\citet{tinney}; (11)~\citet{mnras}; (12)~\citet{weis}.}
\label{known}
\end{deluxetable}

\clearpage
\begin{deluxetable}{ccccrrrrccccccc}
\tabletypesize{\scriptsize}
\rotate
\tablewidth{0pt}
\tablecolumns{15}
\setlength{\tabcolsep}{0.060in}
\tablecaption{Data for Targets Included in NLTT Sample~1.}

\tablehead{
\colhead{NLTT/LP}   &  \colhead{Other}  &  \colhead{$\alpha$}       &   \colhead{$\delta$}            &     \colhead{$m_r$}  &  \colhead{$J$}     &  \colhead{$H$}    &  \colhead{$K_s$}     &  \colhead{$M_J$(TiO 5)}         &  \colhead{$M_J$(CaH 2)} & \colhead{$M_J$(CaOH)}  &  \colhead{$M_J$} &\colhead{d$_{ph}$} & \colhead{Sp.} & \colhead{20}\\
\colhead{Name}   &  \colhead{Name}  &  \colhead{(J2000)}       &   \colhead{}            &     \colhead{}  &  \colhead{}     &  \colhead{}    &  \colhead{}     &  \colhead{}         &  \colhead{} & \colhead{}  &  \colhead{} &\colhead{(pc)} & \colhead{Type} & \colhead{pc?}}

\startdata
405-  5 &             LHS 1060 &                  00 21 16.44 & \phs18 43 56.02 &15.7 &11.352 &10.758 &10.400  &$ 9.86\pm0.19$  &                   $ 9.47\pm0.19$ &                   $ 9.68\pm0.28$ &                 $ 9.67\pm0.20$ &  $ 21.8\pm2.0$ &             M5.0 &  ? \\
645- 53 &               &                         00 35 44.13 &$-$05 41 10.20   &14.9 &10.717 &10.084 & 9.716  &$ 9.65\pm0.19$  &                   $ 9.07\pm0.19$\tablenotemark{a} &  $ 9.39\pm0.28$ &                 $ 9.37\pm0.27$ &  $ 18.8\pm2.3$ &             M5.0 &  Y \\
647- 13 &               &                         01 09 51.17 &$-$03 43 26.41   &17.9 &11.695 &10.921 &10.418  &  \nodata       &                     \nodata      &                     \nodata      &                   \nodata      &  $ 10.5\pm3.0$\tablenotemark{e} & M9.0 &  Y \\
467- 17 &               &                         01 12 00.03 & \phs15 02 17.20 &16.9 &11.928 &11.321 &10.965  &$ 9.89\pm0.19$  &                   $ 9.27\pm0.19$ &                   $ 9.37\pm0.28$ &                 $ 9.54\pm0.30$ &  $ 30.3\pm4.1$ &             M5.5 &  N \\
768-113 &             G 272-43 &                  01 33 58.01 &$-$17 38 23.63   &13.7 & 8.876 & 8.268 & 8.003  &$ 7.63\pm0.35$  &                   $ 7.55\pm0.33$ &                   $ 7.80\pm0.32$\tablenotemark{a} &$ 7.67\pm0.22$ &  $ 17.5\pm1.8$ &             M3.5 &  Y \\
352- 79 &               &                         02 03 28.65 & \phs21 34 16.95 &16.7 &11.634 &11.027 &10.700  &$ 9.72\pm0.19$  &                   $ 9.57\pm0.19$ &                   $ 9.01\pm0.28$\tablenotemark{a} &$ 9.52\pm0.33$ &  $ 26.7\pm4.0$ &             M5.0 &  N \\
\enddata
\tablenotetext{a}{This predicted value of $M_J$ is double-valued and the result closest to the predictions based on other indices was used as our estimate $M_J$.}
\tablenotetext{b}{This predicted value of $M_J$ is double-valued and both predictions are more than 3$\sigma$ from either of the other predictions.  Neither are incorporated in calculating the final value $M_J$.}
\tablenotetext{c}{This value is more than 3$\sigma$ from either of the other predictions and was not used to estimate $M_J$.}
\tablenotetext{d}{LP~31-301 and LP~31-302 form a wide binary system with a separation of 5\arcsec.  
The bandstrength indices measured for the brighter star fall in the ambiguous range, and we quote two distances estimates.
The distance derived for the fainter star, LP~31-302, is consistent with the shorter distance of $14.7\pm1.2$ pc.}
\tablenotetext{e}{See text for explanation of distance estimate.}
\tablenotetext{f}{Also identified by \citet{NN4}.}
\tablecomments{
Column 1 lists the designation from the NLTT catalog.  Identifiers without prefixes are Luyten-Palomar (LP) Survey numbers.
Column 2 lists alternate designations.  
Columns 3 and 4 list the position of the 2MASS source.  
Column 5 lists the $m_r$ given in the NLTT.  
Columns 6-8 list the infrared photometry from 2MASS. 
Columns 9-11 give the predicted $M_J$ based on the spectral indices TiO 5, CaH 2, and CaOH. 
Column 12 gives our final estimate of $M_J$.
Column 13 gives the distance estimates.
Column 14 gives the spectral type.
Column 15 indicates whether the object lies within our distance limit of 20 pc (Y), within 1$\sigma$ of the boundary (?), or beyond the limit (N).
}
\label{yes}
\end{deluxetable}

\clearpage
\begin{deluxetable}{ccccrrrrcccccc}
\tabletypesize{\scriptsize}
\rotate
\tablewidth{0pt}
\tablecolumns{14}
\setlength{\tabcolsep}{0.060in}
\tablecaption{Data for Targets Not Included in NLTT Sample~1.}

\tablehead{
\colhead{NLTT/LP}   &  \colhead{Other}  &  \colhead{$\alpha$}       &   \colhead{$\delta$}            &     \colhead{$m_r$}  &  \colhead{$J$}     &  \colhead{$H$}    &  \colhead{$K_s$}     &  \colhead{$M_{J}$(TiO 5)}         &  \colhead{$M_J$(CaH 2)} & \colhead{$M_J$(CaOH)} &  \colhead{$M_J$} &\colhead{d$_{ph}$} & \colhead{Sp.}\\
\colhead{Name}   &  \colhead{Name}  &  \colhead{(J2000)}       &   \colhead{}            &     \colhead{}  &  \colhead{}     &  \colhead{}    &  \colhead{}     &  \colhead{}         &  \colhead{} & \colhead{}  &  \colhead{} &\colhead{(pc)} & \colhead{Type}
}
\startdata
704- 48 &                 &                   00 06 47.47 &    $-$08 52 35.22   &17.0 &11.970 &11.409 &11.059 &$10.27\pm0.19$ &                  $ 10.15\pm0.19$ &                    $ 10.27\pm0.28$ &                 $ 10.22\pm0.13$ & $ 22.4\pm1.4$ & M6.0 \\  
880-442\tablenotemark{d} & G 267-33 &         00 07 22.50 &    $-$29 35 17.00   &14.8 &11.285 &10.745 &10.518 &$ 8.70\pm0.19$ &                  $  8.50\pm0.19$ &                    $  8.80\pm0.28$ &                 $  8.64\pm0.17$ & $ 33.9\pm2.7$ & M3.5 \\  
  &                       &                               &                     &     &       &       &       &$ 7.72\pm0.35$ &                  $  7.62\pm0.33$ &                    $  7.85\pm0.32$ &                 $  7.73\pm0.21$ & $ 51.6\pm5.0$ &      \\
880-441\tablenotemark{d} &  &                 00 07 22.50 &    $-$29 35 25.67   &18.0 &13.017 &12.481 &12.135 &$10.11\pm0.19$ &                  $  9.64\pm0.19$ &                    $ 10.86\pm0.28$ &                 $ 10.06\pm0.52$ & $ 40.1\pm9.4$ & M5.5 \\  
824-383\tablenotemark{i} &                 &                   00 12 27.48 &    $-$25 28 00.72   &18.9 &13.155 &12.543 &12.092 &$ 9.81\pm0.19$ &                  $ 10.13\pm0.19$ &                    $  5.65\pm0.32$\tablenotemark{c} &$ 9.97 \pm0.21$ & $ 43.6\pm4.2$ & M5.5 \\  
150- 53 &                 &                   00 53 42.45 &     \phs49 42 51.39 &18.8 &13.014 &12.467 &12.039 &$10.21\pm0.19$ &                  $  9.98\pm0.19$ &                    $  9.45\pm0.28$ &                 $  9.97\pm0.34$ & $ 41.0\pm6.4$ & M6.0 \\
\enddata
\tablenotetext{a}{This predicted value of $M_J$ is double-valued and the result closest to the predictions based on other indices was used as our estimate $M_J$.}
\tablenotetext{b}{This predicted value of $M_J$ is double-valued and both predictions are more than 3$\sigma$ from either of the other predictions.  Neither are incorporated in calculating the final value $M_J$.}
\tablenotetext{c}{This value is more than 3$\sigma$ from either of the other predictions and was not used to estimate $M_J$.}
\tablenotetext{d}{LP~880-441 and LP~880-442 form a wide binary system with separation of 8\arcsec.  As with LP~31-301/302, 
the brighter star has indices which fall within the region of ambiguity in Figure~\ref{mag_cal}.  
However, in this case the distance estimate to the fainter star is not sufficiently accurate to provide an 
improved distance estimate to the system.}
\tablenotetext{e}{Has common proper motion with G 4-5.}
\tablenotetext{f}{Has common proper motion with G 36-39.}
\tablenotetext{g}{Has common proper motion with LP~831-38.}
\tablenotetext{h}{Has common proper motion with G 8-48.}
\tablenotetext{i}{Candidate M subdwarf, see \S\S\ref{diskdwarfs} and \ref{sub_dwarf}}
\tablecomments{The columns are the same as the previous table.}
\label{no}
\end{deluxetable}

\clearpage
\begin{deluxetable}{lccc}
\tablewidth{0pt}
\tablecolumns{4}
\tablecaption{Regions That Define the Spectroscopic Indices}

\tablehead{
\colhead{Index}   &  \colhead{Numerator ($\mbox{\AA}$)}  &  \colhead{Denominator ($\mbox{\AA}$)}       &   \colhead{Ref}}
\startdata
CaOH &    6230--6240 &                       6345--6354 &                        1 \\
H$\alpha$ &6560--6566 &                     6545--6555 &                        1 \\
CaH 1 &   6380--6390 &                       Avg. of 6345--6355 and 6410--6430 & 1 \\
CaH 2 &   6814--6846 &                       7042--7046 &                        1 \\
CaH 3 &   6960--6990 &                       7042--7046 &                        1 \\
TiO-a &   7033--7048 &                       7058--7073 &                        2 \\
TiO 2 &   7058--7061 &                       7043--7046 &                        1 \\
TiO 3 &   7092--7097 &                       7079--7084 &                        1 \\
TiO 4 &   7130--7135 &                       7115--7120 &                        1 \\
TiO 5 &   7126--7135 &                       7042--7046 &                        1 \\
VO-a &    Sum of 7350--7370 and 7550--7570 & 7430--7470 &                        2 \\
PC3  &    8230--8270 &                       7540--7580 &                        3 \\ 
\enddata
\tablerefs{(1)~\citet{K99}; (2)~PMSU1; (3)~\citet{ma2}}
\label{defn}
\end{deluxetable}

\clearpage
\begin{deluxetable}{llcccccccccccc}
\tabletypesize{\small}
\rotate
\tablewidth{0pt}
\tablecolumns{14}
\setlength{\tabcolsep}{0.058in}
\tablecaption{Spectral Indices for All Targets and Atandards}

\tablehead{
\colhead{Name}&  \colhead{2MASS}&  \colhead{CaOH}&   \colhead{H$\alpha$} &  \colhead{CaH 1} &  \colhead{CaH 2} &  \colhead{CaH 3} &  \colhead{TiO-a} & \colhead{TiO 2}  &  \colhead{TiO 3} &\colhead{TiO 4} & \colhead{TiO 5} & \colhead{VO-a} & \colhead{PC3} \\
\colhead{}  &  \colhead{Name} &  \colhead{}    &   \colhead{}           & \colhead{}      &  \colhead{}     &  \colhead{}      &  \colhead{}      & \colhead{}       &  \colhead{}      &\colhead{}      & \colhead{}  & \colhead{} & \colhead{}}
\startdata
LP 704- 48 & 2MASSI0006$-$0852      &        0.209 &   0.952 & 0.793 & 0.252 & 0.566 & 2.397 & 0.334 & 0.614 & 0.462 & 0.209 &2.052   &    1.331\\
LP 880-442 & 2MASSI0007225$-$293526 &        0.411 &   0.933 & 0.807 & 0.429 & 0.686 & 1.484 & 0.616 & 0.705 & 0.663 & 0.421 &1.917   &    0.912\\
LP 880-441 & 2MASSI0007225$-$293517 &        0.129 &   0.906 & 0.826 & 0.307 & 0.588 & 2.113 & 0.419 & 0.605 & 0.476 & 0.230 &2.012   &    1.210\\
LP 824-383 & 2MASSI0012$-$2528      &        0.884 &   1.512 & 1.146 & 0.255 & 0.444 & 2.970 & 0.238 & 0.545 & 1.192 & 0.272 &2.075   &    1.257\\
LP 292- 67 & 2MASSI0020+3305        &        0.316 &   0.942 & 0.787 & 0.333 & 0.654 & 2.123 & 0.416 & 0.555 & 0.542 & 0.251 &2.070   &    1.207\\
LP 405-  5 & 2MASSI0021+1843        &        0.290 &   1.106 & 0.783 & 0.325 & 0.620 & 1.915 & 0.452 & 0.591 & 0.506 & 0.264 &2.012   &    1.214\\
LP 645- 53 & 2MASSI0035$-$0541      &        0.330 &   0.989 & 0.858 & 0.368 & 0.685 & 1.873 & 0.468 & 0.621 & 0.542 & 0.292 &2.025   &    1.106\\
LP 150- 53 & 2MASSI0053+4942        &        0.321 &   1.336 & 0.783 & 0.271 & 0.547 & 2.643 & 0.328 & 0.511 & 0.592 & 0.217 &2.106   &    1.383\\
\enddata
\label{indicies}
\end{deluxetable}

\clearpage
\begin{deluxetable}{llcccc}
\tabletypesize{\footnotesize}
\tablewidth{0pt}
\tablecolumns{6}
\tablecaption{Activity}

\tablehead{
\colhead{Name} &  \colhead{2MASS} & \colhead{H$\alpha$ flux}                           &  \colhead{H$\alpha$}        & $m_{bol}$ & $\log{F_\alpha \over F_{bol}}$ \\
\colhead{}     &  \colhead{Name}  & \colhead{$10^{-15}\mbox{ergs }cm^{-2}sec^{-1}$}  &  \colhead{EW ($\mbox{\AA}$)} & \colhead{}& \colhead{}}
\startdata
LP 405-  5 & 2MASSI0021+1843        &       5.51 &      2.3  & 13.27 & $-$4.36\\
LP 150- 53 & 2MASSI0053+4942        &       0.85 &      4.2  & 14.95 & $-$4.50\\
LP 150- 58 & 2MASSI0100+5055        &       4.58 &      6.4  & 14.41 & $-$3.99\\
LP 706- 88 & 2MASSI0103$-$1126      &       2.67 &      8.2  & 14.53 & $-$4.17\\
LP 647- 13 & 2MASSI0109$-$0343      &       2.75 &     10.5  & 13.60 & $-$4.53\\
LP 194- 35 & 2MASSI0111+4127        &       4.98 &      9.0  & 14.14 & $-$4.06\\
LP 768-113 & 2MASSI0133$-$1738      &     249.00 &      2.0  & 10.70 & $-$3.74\\
LP 352- 79 & 2MASSI0203+2134        &      61.80 &     26.0  & 13.54 & $-$3.21\\
\enddata
\label{halpha}
\end{deluxetable}


\clearpage

\clearpage 


\clearpage

\clearpage

\clearpage

\clearpage

\clearpage

\clearpage

\clearpage

\end{document}